# Advanced Persistent Threats (APT) Attribution Using Deep Reinforcement Learning

Animesh Singh Basnet and Mohamed Chahine Ghanem *, Dipo Dunsin, Wiktor Sowinski-Mydlarz

*Abstract*—This paper investigates the application of Deep Reinforcement Learning (DRL) for attributing malware to specific Advanced Persistent Threat (APT) groups through detailed behavioural analysis. By analysing over 3,500 malware samples from 12 distinct APT groups, the study utilises sophisticated tools like Cuckoo Sandbox to extract behavioural data, providing a deep insight into the operational patterns of malware. The research demonstrates that the DRL model significantly outperforms traditional machine learning approaches such as SGD, SVC, KNN, MLP, and Decision Tree Classifiers, achieving an impressive test accuracy of 89.27%. It highlights the model's capability to adeptly manage complex, variable, and elusive malware attributes. Furthermore, the paper discusses the considerable computational resources and extensive data dependencies required for deploying these advanced AI models in cybersecurity frameworks. Future research is directed towards enhancing the efficiency of DRL models, expanding the diversity of the datasets, addressing ethical concerns, and leveraging Large Language Models (LLMs) to refine reward mechanisms and optimise the DRL framework. By showcasing the transformative potential of DRL in malware attribution, this research advocates for a responsible and balanced approach to AI integration, with the goal of advancing cybersecurity through more adaptable, accurate, and robust systems.

## I. INTRODUCTION

In recent years, cyber attacks have evolved from isolated incidents into sophisticated operations conducted by well-resourced Advanced Persistent Threats (APTs), which are characterised by their strategic, long-term approaches compared to more opportunistic cyberattacks [1]. Like traditional cyberattacks, APTs utilise malware as their primary tool, but they stand out due to their complexity, higher number of network events, and intricate behavioural activities [2]. APTs are meticulously orchestrated, employing advanced techniques to remain hidden while they extract data, disrupt operations, or create entry points for future attacks [3]. Often backed by nation-states or large organisations with political or economic motives, APTs pose significant threats to critical resources

* Mohamed Chahine Ghanem is the corresponding author.
A. S. Basnet is with Cyber Security Research Centre, London Metropolitan University, London, UK e-mail:anb1380@my.londonmet.ac.ukk
M. C. Ghanem is with the Cybersecurity Institute, University of Liverpool and Cyber Security Research Centre, London Metropolitan University, UK e-mail: mcghanem@liverpool.ac.uk
D. Dunsin is with Cyber Security Research Centre, London Metropolitan University, London, UK e-mail: d.dunsin@londonmet.ac.uk
W. Sowinski-Mydlarz is with Cyber Security Research Centre, London Metropolitan University, London, UK e-mail:w.sowinskimydlarz@londonmet.ac.uk

[1]. A notable example is the Stuxnet worm, which, although discovered in 2010, had been operating covertly since at least 2005, specifically targeting Iran's nuclear facilities at the Natanz uranium enrichment plant [4]. Developed by the USA, Stuxnet utilised advanced evasion techniques like zero-day exploits and rootkits to infiltrate and compromise its target while remaining undetected for years [4].

According to Statista, the global revenue from the APT protection market is worked to reach $12.5 billion by 2025, driven by the urgent need to defend against these evolving threats [5]. Despite significant investments in security solutions, APT incidents, including ransomware attacks, continue to rise across industries, military sectors, and government institutions, with a 55.5% increase in ransomware cases in 2023 alone, reaching 4,368 incidents worldwide [6], [7]. The use of advanced technologies like large language models (LLMs) has further intensified the threat landscape, enabling more sophisticated cyberattacks [8]. This escalation underscores the critical need for innovative defence strategies, encouraging organisations and governments to continuously invest in advanced security measures to stay ahead of these persistent adversaries [9].

The increasing sophistication and frequency of APTs highlight the critical challenge of precise attribution in the cybersecurity landscape [2]. Accurate attribution is essential for developing targeted defensive strategies, as understanding an adversary's tactics, techniques, and procedures (TTPs) allows for tailored responses to specific threats [10]. It also plays a key role in holding perpetrators accountable, which can act as a deterrent through legal and diplomatic consequences, thereby maintaining global cyber stability [11]. However, attribution is complicated by the obfuscation methods used by APTs, including routing attacks through proxies and deceptive indicators [12]. These sophisticated tactics require extensive technical expertise and collaboration across sectors to analyse threat profiles that reveal attackers' motives and strategies [12]. Building on this complexity, each APT group possesses a distinct signature, merging specific malware applications with strategic objectives, whether for financial gain or disrupting critical infrastructure [5]. The intricacies of these profiles underscore the importance of attribution, pinpointing the perpetrator not only aids in defence but also in shaping cybersecurity policies and measures to pre-empt future attacks [13].

To address the growing challenge of attributing APTs, this report suggests leveraging machine learning algorithms that focus on analysing malware behaviour within sandbox environments. Machine learning's ability to process vast datasets and detect subtle patterns offers a promising solution to understanding the complex and often hidden techniques used by APT groups [10]. By training models on behavioural data obtained from executing malware in virtual systems, these systems can be developed to automatically detect and classify patterns, leading to more precise attribution of cyberattacks. Within this context, Deep Reinforcement Learning (DRL) emerges as particularly effective for attributing APT malware. DRL combines deep learning's pattern recognition capabilities with reinforcement learning's adaptive decision-making through trial and error, enabling it to detect and enhance its response to evolving malware behaviours [14]. Techniques like the Markov Decision Process (MDP) and model-free learning allow DRL to structure decision-making and adapt without relying on predefined models. Unlike traditional machine learning models that may struggle with the dynamic nature of cyber threats, DRL continuously learns and refines its strategies, making it highly effective against sophisticated APT tactics. Its ability to operate in environments with incomplete information, simulate diverse attack scenarios, and evolve through interaction underscores its potential as a powerful tool in crafting robust cyber defence strategies [15].

## II. RELATED WORK

The adoption of Deep Reinforcement Learning (DRL) for malware detection is a relatively recent and promising development in the wider field of cybersecurity [16]. Grasping the related work in this area requires an initial understanding of the behavioural patterns, origins, and classifications of malware. Since malware plays a crucial role in APT attacks, analysing the characteristics of these malware attacks can reveal essential attributes of the attacking APT group [17]. The chosen papers examine current research and methodologies in malware behavioural analysis, attribution, and family classification, critically assessing the effect of these elements to enhance the precision and efficiency of cybersecurity defences [18].

### A. Behavioural Analysis

Malware behavioural analysis is a cornerstone technique in cybersecurity that involves observing and understanding the actions performed by malware within a controlled environment, typically a sandbox [19]. This technique allows for the identification of malicious patterns and behaviours including networks and operations within the system. Recent studies emphasise the evolution of this analysis to include automated systems that leverage machine learning to predict and react to malware behaviour dynamically [20]. Such systems can discern between benign and malicious processes by examining changes made by the software to the system's state or its network behaviour [21]. These analyses often involve the extraction of features such as API calls, file-system operations, and network activity which are then processed using advanced algorithms to detect anomalous patterns that suggest malicious intent [21]. By comprehensively understanding the behaviour exhibited by the malware during execution and examining its underlying code and structure, we can gain valuable insights that aid in accurately attributing the malware to specific APT groups or threat actors [22].

### B. Malware Attribution

The attribution of malware, identifying the probable origin or actor behind an attack, is a complex yet crucial task within cybersecurity. Traditional approaches in malware attribution have relied heavily on manual, domain-specific feature engineering and pre-processing to isolate attributes indicative of a malware's lineage or family ties [23]. The incorporation of neural networks has significantly advanced malware attribution capabilities, particularly using machine learning techniques such as Random Forest and Extreme Gradient Boosting (XG-Boost), which have strengthened efforts in this area [24]. These algorithms refine neural network models by improving accuracy and handling overfitting, essential for distinguishing between benign and malicious activities in vast and complex datasets [24]. This synergy optimises feature selection and enhances predictive capabilities, effectively supporting the identification of malware origins and behaviours.

Recent studies have introduced novel approaches in malware attribution that significantly improve upon traditional methods. For instance, the work by Binhui Tang and colleagues transforms APT malware samples into RGB images rather than relying on standard grayscale feature extraction [25]. This approach allows for deeper and more nuanced feature mining, using an enhanced Convolutional Neural Network (CNN) model that incorporates Self-Attention mechanisms and Spatial Pyramid Pooling (SPP-net) [25]. This novel framework aids in not only detecting APT malware but also in facilitating the identification of malware origins and attack methodologies through sophisticated visual data representations. Another innovative approach is presented by Elijah Snow and his team, who utilised an end-to-end multimodal learning strategy [26]. This method integrates three distinct neural network architectures—dense networks, CNNs, and Recurrent Neural Networks (RNNs) with Long Short-Term Memory (LSTM) cells—to automatically extract and learn features from diverse malware data attributes [26]. By combining these architectures, their model effectively classifies malware into respective groups, enhancing the granularity and accuracy of malware attribution. Further, Gil Shenderovitz and Nir Nissim introduced a dynamic analysis technique for segmenting Multivariate Time Series Data (MTSD) derived from API calls [27]. Their approach uses temporal segmentation to provide a detailed behavioural profile of APT malware, facilitating the detection and attribution to specific cyber-groups or nations with enhanced explainability [28]. The work of Jian Zhang and colleagues improved the integration of multiple feature dimensions by employing a Graph Neural Network (GNN) model to create an event behaviour graph based on API instructions and operations, combined with an innovative ImageCNTM for capturing local spatial correlations and long-term dependencies of opcode images [29]. By fusing word frequency and behavioural features in a multi-input deep learning model, they propose a comprehensive system that classifies and accurately attributes APT malware, improving upon traditional single-dimensional models. Similarly, Shudong Li and his team refined the classification methodology by implementing a dynamic analysis and preprocessing stage for malware samples, followed by feature representation using the TF-IDF method and feature dimensionality reduction using the chi-square test [30]. Their innovative use of a Multiclass SMOTE-RF model addresses class imbalance, enhancing the classification accuracy significantly across various malware families [30].

However, despite these advancements, challenges persist, particularly when dealing with malware that lacks evolutionary links or belongs to completely different families. Traditional methods, as noted by Rosenberg and colleagues, often fall short in such scenarios because they primarily focus on detecting mutations or similarities within the same functional group [23]. This limitation highlights the need for more advanced and flexible analytical tools capable of handling a broad spectrum of malware types, moving beyond familial or evolutionary similarities to embrace a more holistic and integrative approach in malware analysis [31].

Addressing these challenges, this work presents a novel approach utilising Deep Reinforcement Learning (DRL) for malware attribution, specifically tailored for APTs developed by nation-states. DRL has previously shown significant advancements in malware detection, effectively identifying and responding to APT activities. For instance, Cho Do Xuan and Nguyen Hoa Cuong have developed the FIERL model, which employs BiLSTM and Attention networks to extract unusual behaviour from network traffic data involving APT and normal IPs, further enhancing detection capabilities through data rebalancing and contrastive learning for APT IP classification [32]. Similarly, Kazeem Saheed and Shagufta Henna applied DRL to wireless network traffic data, where the system dynamically learns and adapts to new APT attack strategies, showcasing an ability to outperform traditional models by rapidly adjusting to evolving threats [6]. Additionally, Mangadevi Atti and Manas Kumar Yogi utilised a DRL framework that leverages Proximal Policy Optimization (PPO) to learn complex patterns from executable files, optimising malware detection processes and enhancing the system's predictive accuracy [16]. These applications underscore DRL's pivotal role in the detection domain, demonstrating its potential not only for identifying malware but also for attributing it effectively to specific APTs developed by nation-states.

DRL's application in the context of malware attribution offers a significant advancement over previous methods, as it does not rely on pre-defined models or static features, which are often limited by the need for extensive manual extraction and are less effective across disparate malware families [33]. DRL leverages the strengths of deep learning for pattern recognition within complex and large-scale datasets, combined with the strategic decision-making capabilities of reinforcement learning [34]. This approach is particularly adept at processing incomplete or obfuscated data commonly employed in sophisticated cyberattacks, enabling it to adaptively learn and predict attribution based on behavioural patterns rather than static signatures.

## III. RESEARCH QUESTIONS AND CONTRIBUTION

As previous sections have detailed the complexity and threat posed by APTs, this study leverages the sophisticated capabilities of DRL to analyse and interpret intricate malware data from controlled tests. The main goal is to refine a DRL model that effectively attributes APTs by analysing behavioural patterns, thus advancing cybersecurity defence mechanisms. This initiative to apply DRL seeks to harness its superior pattern recognition and strategic decision-making properties to enhance the detection and mitigation of advanced cyber threats.

### A. Research Questions

The guiding questions of this research aim to critically evaluate the effectiveness of DRL in the cybersecurity landscape, particularly in attributing APTs. These questions explore: the identification of unique behavioural patterns of APTs within sandbox-analysed malware, the capability of DRL to precisely differentiate between malware behaviours from diverse APT groups, and the influence of the Markov Decision Process in boosting the strategic decision-making of DRL models within the context of cyber threat attribution. These inquiries are designed to assess whether DRL can offer a sophisticated and adaptive approach to understanding and countering APTs.

### B. Contribution

This study makes impactful contributions to the domain of cybersecurity by pioneering the use of Deep Reinforcement Learning (DRL) for the specific purpose of APT attribution, benchmarking its effectiveness against traditional machine learning models, and exploring its adaptability to varied APT scenarios. It constructs a DRL model that not only processes and understands detailed behavioural data from malware but also empirically demonstrates its enhanced effectiveness over existing techniques. Additionally, by probing the model's ability to adapt to new threats, the research highlights DRL's potential to evolve and maintain relevance in a rapidly changing threat environment. The findings and methodologies of this research expand the practical and theoretical frameworks for deploying advanced AI in active cybersecurity defences, potentially setting new standards for the integration of machine learning in threat intelligence and response strategies.

## IV. METHODOLOGY

This section outlines the methodology adopted during the design, implementation and testing of the system. it provides details and justification on tools, approaches and methods employed as well as providing background information necessary for understanding the methodology.

### A. Proposed System Design

The research adopts an experimental and simulation-based design, focusing on developing and evaluating a Deep Reinforcement Learning model for malware attribution to APT groups. The study begins by preparing a dataset of malware samples, followed by data preprocessing and feature extraction to ensure accuracy and relevance. The DRL model is then trained and tested in a simulated environment designed to mimic real-world conditions. This allows for controlled experimentation, where the model's ability to handle complex and evasive malware behaviours can be systematically assessed using metrics such as accuracy, robustness, and computational efficiency.

### B. Data Collection

Data collection is strategically executed from two specialised sources to capture a broad spectrum of malware behaviours and characteristics, ensuring the depth and breadth of data necessary for effective DRL modelling.

*1) Cuckoo Report:* The Cuckoo Sandbox is an advanced open-source malware analysis system designed to analyse and report on the behaviour of potentially malicious files in a secure, isolated environment [35], [36]. It is widely used for malware detection by providing a controlled setting where files can be executed to observe their actions without risking the integrity of the host system [35]. In this work, the Cuckoo Sandbox plays an integral role in the hybrid process of collecting malware behavioural data, combining both manual and automated tasks to efficiently analyse large datasets of malware samples.

Specifically, because this analysis utilises the web version of Cuckoo Sandbox, mallicious files are manually uploaded via the Cuckoo Sandbox interface, which provides detailed system specifications and analytics, then prioritise and monitor these files in a secure environment to ensure the host system's integrity. Process IDs are then generated during the analysis phase, which allow tracking and retrieval of detailed reports on each malware sample's activity, including system changes and network traffic. An automated script further facilitates the extraction of these IDs, organising them into a structured dataset and performing data cleanup to avoid duplicates, ensuring the integrity and accuracy of the analysis.

*2) VirusTotal Report:* VirusTotal is a comprehensive online service that analyses files and URLs to detect viruses, worms, trojans, and other kinds of malicious content [37], [38]. Leveraged by security professionals and researchers, VirusTotal aggregates information from over 70 antivirus scanners and URL/domain blacklisting services, along with a plethora of tools for the analysis of files, which makes it an indispensable resource for the real-time detection of emerging threats [37].

In this work, VirusTotal complements the Cuckoo Sandbox by offering an extensive database of antivirus scan results and behaviour reports for more in-depth analysis of malware samples. Automated scripts access specific API endpoints using the SHA-256 hash of each sample to fetch detailed file reports and behaviour summaries. The file report encapsulates antivirus detection results, file type, size, and associated detection names, while the behaviour report sheds light on the malware's activities within an operating system, such as registry modifications and network actions. These reports are critical for understanding the malware's potential impact and aid in training deep reinforcement learning models to recognise and predict similar behaviours in future security threats.

### C. Data Understanding

Understanding the data collected from sources like Cuckoo Sandbox and VirusTotal is essential before diving into deeper analyses or model development, as it establishes the groundwork for recognising patterns, anomalies, and intrinsic properties of malware behaviours. This preliminary step ensures that subsequent processes, such as data cleaning, preprocessing, and detailed exploratory analysis, are effectively tailored to the characteristics of the data. For instance, the "reports.json" file from Cuckoo Sandbox provides a wealth of information on malware activities through detailed logs of file creation, registry changes, and network connections. By parsing these entries, it is possible to discern the common tactics used by malware, such as communication strategies and system infiltration methods, which are crucial for identifying threat behaviours.

Similarly, VirusTotal's file and behaviour reports complement the data by providing insights into the malware's detectable characteristics and operational tactics within infected systems. These reports include critical metadata on the malware's type, the extent of its recognition across different security platforms (unique_sources), and its evasion techniques (packers). Additionally, behavioural data like registry modifications and network traffic from these reports help in understanding how malware interacts with and affects systems, highlighting potential persistence mechanisms or damage attempts. Through a comprehensive understanding of these datasets, it is possible to understand that the data used in modelling is accurate, reliable, and robust enough to develop effective machine-learning models that can attribute malware to specific APT groups, enhancing cybersecurity measures and threat intelligence.

### D. Data Preparation

Data preparation is crucial for transforming raw data from Cuckoo Sandbox and VirusTotal into a structured format suitable for in-depth analysis and modelling. The process begins with data cleaning, which involves refining the datasets to highlight essential malware characteristics. For file reports, this includes isolating key attributes like "file_name", "apt_group", and "unique_sources", and quantifying the threat level by analysing entries classified as "malicious". Additionally, the "import_list" is parsed to assess the complexity of malware interactions. For behaviour reports, the focus is on dynamic interactions, such as the number of files written and registry keys manipulated, which provide insights into the malware's impact on system operations. Cuckoo reports are also processed to extract API call statistics from "api_stats", giving a detailed view of system interactions at the API level.

Following data cleaning, the process moves to data integration, where the cleaned datasets are merged into a cohesive framework for unified analysis. The file and behaviour datasets are merged with the Cuckoo reports, with missing entries filled with zeros to maintain numerical data integrity. This step is essential for creating a comprehensive dataset that aligns all aspects of the malware's behaviour, enabling more effective modelling and analysis.

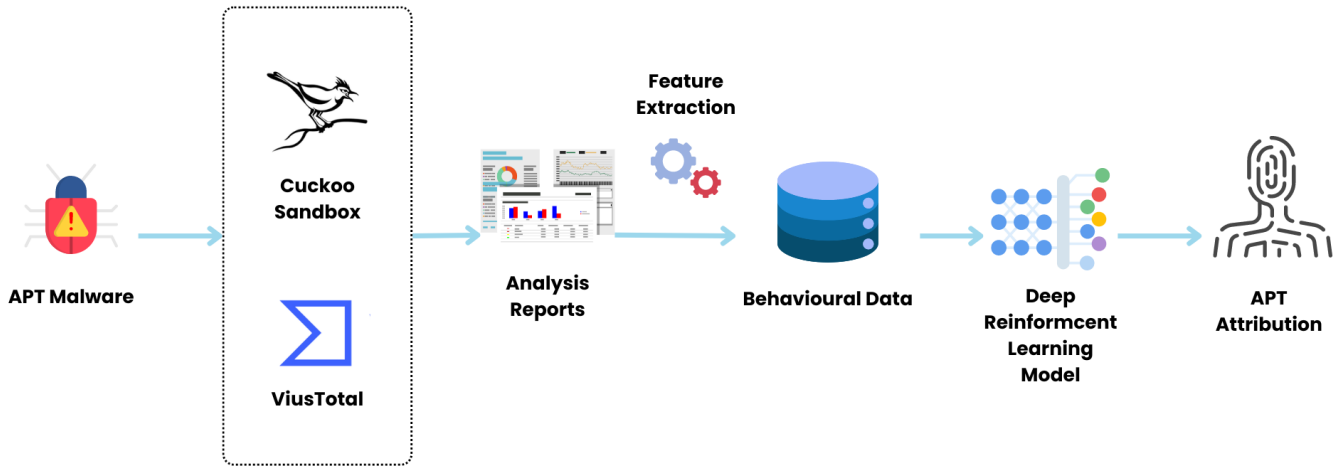

Fig. 1: System workflow for the proposed APT Attribution Model

*E. Data Modelling*

In the data modelling phase, several critical steps are undertaken to prepare the dataset for effective machine learning applications. The process begins with Variable Transformation, where the dataset variables are identified and categorised based on their data types. Numerical columns are separated from categorical columns to facilitate different preprocessing techniques suitable for each type. The "apt_group" column, serving as the target variable for the models, is meticulously handled to ensure it is excluded from the feature sets when present in numerical columns, preventing data leakage. Categorical variables are then transformed into integer codes using techniques like Label Encoding, converting nominal data into a format that is digestible for machine learning algorithms. This transformation is essential for preparing the data for accurate and efficient modelling, ensuring that all features are in a machine-readable form.

Following the transformation, the dataset undergoes Data Partition and Class Imbalance treatment and normalisation to optimise it for model training and evaluation. SMOTE (Synthetic Minority Over-sampling Technique) is employed to address class imbalances within the dataset, synthesising new examples in the minority class to prevent model bias towards the majority class. This ensures a balanced representation across classes, which is crucial for generalising the model effectively. The data is then split into training and testing sets, with a significant portion reserved for testing to assess model performance. Subsequently, normalisation is performed using MinMaxScaler, scaling all features to a uniform range to prevent any single variable from dominating due to its scale.

This step is vital as it allows the machine learning model to converge more rapidly during training. The normalised data is then carefully reformatted back into DataFrames, retaining the original column names for better traceability and clarity during model training and evaluation phases.

*F. Model Building*

In the model-building phase, a bespoke environment is crafted using the Gymnasium framework, tailored to the complexities of Advanced Persistent Threat (APT) data. This setup precisely defines the observation space based on the feature set derived from malware samples and the action space aligned with unique labels constructed using the number of APT groups. The environment facilitates the simulation of interaction sequences, rewarding the model for accurate predictions and resetting for new episodes as data points are iteratively processed [39].

During the training of the model, a Deep Q-Network (DQN) is utilised, configured with adjustable learning rates and buffer sizes to optimise the learning curve. The model's performance is periodically assessed using key metrics such as accuracy and the F1 score, which aid in fine-tuning the training regimen. This dynamic approach ensures a balance between exploration of new strategies and exploitation of known effective tactics, enhancing the model's ability to make progressively more accurate malware classifications. Subsequent development includes hyperparameter tuning—adjusting the discount factor, exploration rate (epsilon), and mini-batch sizes—to enhance the learning process's efficiency and effectiveness. Training episodes are varied in length to reflect the complex nature of real-world APT scenarios better, preventing overfitting and improving generalisation. Moreover, regularization techniques like dropout and batch normalisation are integrated within the neural network architecture to mitigate the risk of overfitting by moderating less predictive features' influence and stabilising learning across different batches. Detailed performance analysis and error metrics are continuously collected and reviewed to identify the model's strengths and weaknesses, providing a clear direction for its ability to capture the APT groups.

### G. Dataset

The APT Malware Dataset utilised in this work is a comprehensive collection of over 3,500 malware samples (https://github.com/cyber-research/APTMalware), categorised into 12 distinct Advanced Persistent Threat (APT) groups obtained from. These groups are believed to be state-sponsored by five different countries, including China, Russia, North Korea, the USA, and Pakistan. The dataset serves as a critical resource for benchmarking various machine-learning techniques aimed at authorship attribution of cyberattacks. [40]

TABLE III: APT Malware Dataset Distribution [40]

| Country | APT Group | Sample Size |
|---|---|---|
| China | APT 1 | 405 |
| China | APT 10 | 244 |
| China | APT 19 | 32 |
| China | APT 21 | 106 |
| Russia | APT 28 | 214 |
| Russia | APT 29 | 281 |
| China | APT 30 | 164 |
| North-Korea | DarkHotel | 273 |
| Russia | Energetic Bear | 132 |
| USA | Equation Group | 395 |
| Pakistan | Gorgon Group | 961 |
| China | Winnti | 387 |
| **Total** |  | **3594** |

Each APT group in the APT Malware Dataset represents a unique threat actor with a specific set of malware samples attributed to their cyber activities. The dataset's diversity is evident in both the quantity of samples per group and the variety of file types, ranging from executable files like .dll and .exe to documents such as .doc, .xlsx, and .ppt. This assortment adds complexity to the analysis, enabling robust evaluations of the various attack vectors and infection methods used by these groups. However, the dataset also shows a significant class imbalance, with some groups having as few as 32 samples and others as many as 961, presenting a challenge for reinforcement learning models to achieve unbiased behavioural representations. To manage this, the samples are meticulously labeled with their SHA-256 hashes for precise identification and stored in separate, password-protected compressed folders to ensure security and data integrity, with the universal password "infected" providing controlled access. The analysis leverages tools like Cuckoo for dynamic analysis, where malware files are extracted and executed, and VirusTotal, which uses the hashes to fetch pre-generated reports for deeper insights into the malware behaviour.

### H. MDP Model

The Markov Decision Process (MDP) provides a structured framework for understanding how an agent makes decisions while interacting with its environment [41]. In this work, MDP framework is utilised to design and develop the DRL model for attributing malware to APT groups. The primary data sources for the model come from detailed reports generated by Cuckoo Sandbox and VirusTotal, which offer comprehensive behavioural analyses of malware samples. These reports provide a multi-dimensional view of each malware's characteristics and behaviour, which are crucial for defining the states, actions, and rewards in the MDP framework as listed below:

*1) States Space:* The state represents the current understanding of a malware sample based on its observed behaviours and characteristics. Each state is derived from a feature dataset that encapsulates various aspects of malware behaviour, such as file operations, registry changes, network activities, and other dynamic interactions recorded during the malware's execution. This dataset is constructed using key data points extracted from the Cuckoo and VirusTotal report. These features collectively form a comprehensive behavioural profile of the malware, encapsulating its operational tactics and techniques, which are used to define the current state in the MDP. This state representation serves as the foundation for the reinforcement learning model's decision-making process, enabling accurate attribution and classification of malware to specific APT groups.

*2) Actions Space:* In the MDP model, an action refers to the transition from analysing one malware sample to another within the dataset. Each action involves selecting a new malware sample from the dataset and performing the analysis to obtain its behavioural profile, thus transitioning the state of the MDP from the current malware profile to the next. This action reflects the decision-making process in identifying and comparing malware attributes across different samples, which is central to attributing them to specific APT groups.

*3) Rewards:* The reward in our proposed MDP model is defined by the accuracy of the attribution. When the DRL model correctly attributes a malware sample to its respective APT group based on the analysed behaviours, a positive reward is assigned. Conversely, incorrect attributions yield a negative reward. The magnitude of the reward is scaled based on the confidence level of the attribution and the criticality of correctly identifying specific APT-related malware, reflecting the importance of precision in cybersecurity measures.

In summary, the dataset features, such as process call count, registry access patterns, file operations, and network behaviours, represent the various states of malware samples within the MDP framework. The DRL agent continuously monitors these states and takes actions to attribute the malware to a specific APT group. At each time step $t$, the agent is in the state $s_t$ and selects an action $a_t$, transitioning to a new state $s_{t+1}$, which corresponds to the analysis of another set of behavioural features. The agent is rewarded based on the accuracy of its attributions, receiving a positive reward (+1) for correct classifications and no reward (0) for misclassifications. Through this process, the agent refines its policy, improving its ability to attribute malware samples to the correct APT group as it progresses through the dataset.

## V. IMPLEMENTATION AND TESTING

### A. Simulation Environment

The proposed DRL model for Advanced Persistent Threat (APT) attribution utilises a structured approach incorporating an environment for sequential decision-making, a Q-network for estimating the quality of actions, and a replay memory for learning from past experiences. Here, the agent's states are derived from comprehensive behavioural data extracted from malware reports, while actions represent decisions to attribute malware to specific APT groups. An action represented by $a_t = n$ indicates the model's prediction, where $n$ corresponds to the malware associated with an APT group. The agent operates within this environment, aiming to optimise the cumulative rewards over time, where the rewards are aligned with the accuracy of the attribution to the correct APT group. This structure is designed to refine the agent's decision-making process and improve its policy through continuous learning and adaptation based on detailed malware behaviour analysis.

*1) Environment:* This is the simulated setting where the DRL agent is deployed, designed for making informed attributions of malware to specific Advanced Persistent Threat (APT) groups based on behavioural analysis. This environment is an adaptation of the OpenAI Gym interface, featuring a discrete action space that corresponds to different APT groups identified in the dataset [42]. The observation space is constructed from detailed features such as API calls, file-system operations, and network activities, which are crucial for defining the states of the malware being analysed [42].

*2) Q-Network:* At the core of the decision-making process, the Q-Network includes a policy network and a target network, each configured as a multilayer perceptron with two hidden layers leading to an output layer that represents each potential APT group. The networks use Leaky ReLU activation functions to maintain gradient flow during training, helping to prevent the vanishing gradient problem that can occur with standard ReLU functions if negative values are present in the inputs [43]. The output layers of the networks apply a MinMaxScaler to normalise the outputs, ensuring that the classification probabilities for the APT groups are scaled between 0 and 1 [44]. This normalisation helps stabilize the learning process by keeping the network's predictions within a consistent range.

*3) Replay Memory:* Essential for robust learning, Replay Memory archives tuples of the agent's experiences, including states, actions, rewards, and subsequent states. These experiences are accumulated as the agent processes the behavioural data, employing an epsilon-greedy strategy $\epsilon$ to balance the exploration of new strategies with the exploitation of known patterns. Each action—representing an attribution decision—transitions the agent from one state to another ($s_t$ to $s_{t+1}$), with rewards assigned based on the accuracy of these attributions.

*4) Policy Training:* The training of the DRL agent's policy operates over a series of episodes, with each episode consisting of numerous time steps, labelled as $T$. Each time step $t$ involves the sampling of a feature vector representing the current state $s_t$ from the replay buffer $\mathcal{B}$, which is then fed into the policy network. The policy network processes this input to output Q-values, $Q(s_t, a_t)$, for potential actions aimed at matching these values with the target or optimal Q-value, $Q(s, a)$.

Once the training process, detailed above, is complete, the efficacy of the agent's policy is evaluated by deploying the policy network model in a test environment. This test environment is carefully constructed using the validation dataset, allowing for a thorough assessment of the model ability to perform under conditions that simulate real-world scenarios.

### B. Experimental Specifications

The experimental setup for the DRL-based APT detection model involves finely tuned parameters for the Deep Q-Network (DQN) model to optimise its performance in learning and adapting to detect advanced persistent threats (APTs) effectively. This section details the specific arguments and configurations passed to the DQN model, which are crucial in defining its learning behaviour and operational dynamics in the simulated environment.

### C. Software Environment

In order to implement and evaluate the DRL-based APT attribution model, several key software tools and techniques are utilised. These are essential for creating a robust environment that can simulate real-world scenarios and evaluate the performance of the model under controlled conditions.

*1) Cuckoo Sandbox:* Cuckoo Sandbox is an open-source automated malware analysis system that acts as a vital tool in the environment. It allows for the isolation and analysis of suspicious files in a safe, contained environment. This sandboxing technique enables the collection of detailed analysis about the behaviour of the file while running in an operating system, which is vital for training the DRL model to recognise threat behaviours. The outputs provided by Cuckoo Sandbox include API calls, network traffic, file system changes, and memory dumps, which serve as critical inputs for the model's learning process. [45]

*2) Stable Baselines 3:* Stable Baselines 3, an enhancement over the original OpenAI Baselines, offers refined implementations of reinforcement learning algorithms. The Deep Q-Network (DQN) model from Stable Baselines 3 is specifically utilised for the agent's training process. This model efficiently estimates the optimal action-value function, which is central to making informed decisions in the simulated network environments. The DQN supports the development of a robust policy that can effectively identify and differentiate between benign and malicious network traffic. [46]

*3) Gymnasium:* Gymnasium, formerly known as Gym, is a tool from OpenAI that provides standardised interfaces for a diverse array of environments. These environments serve as testbeds for reinforcement learning algorithms. In this work,

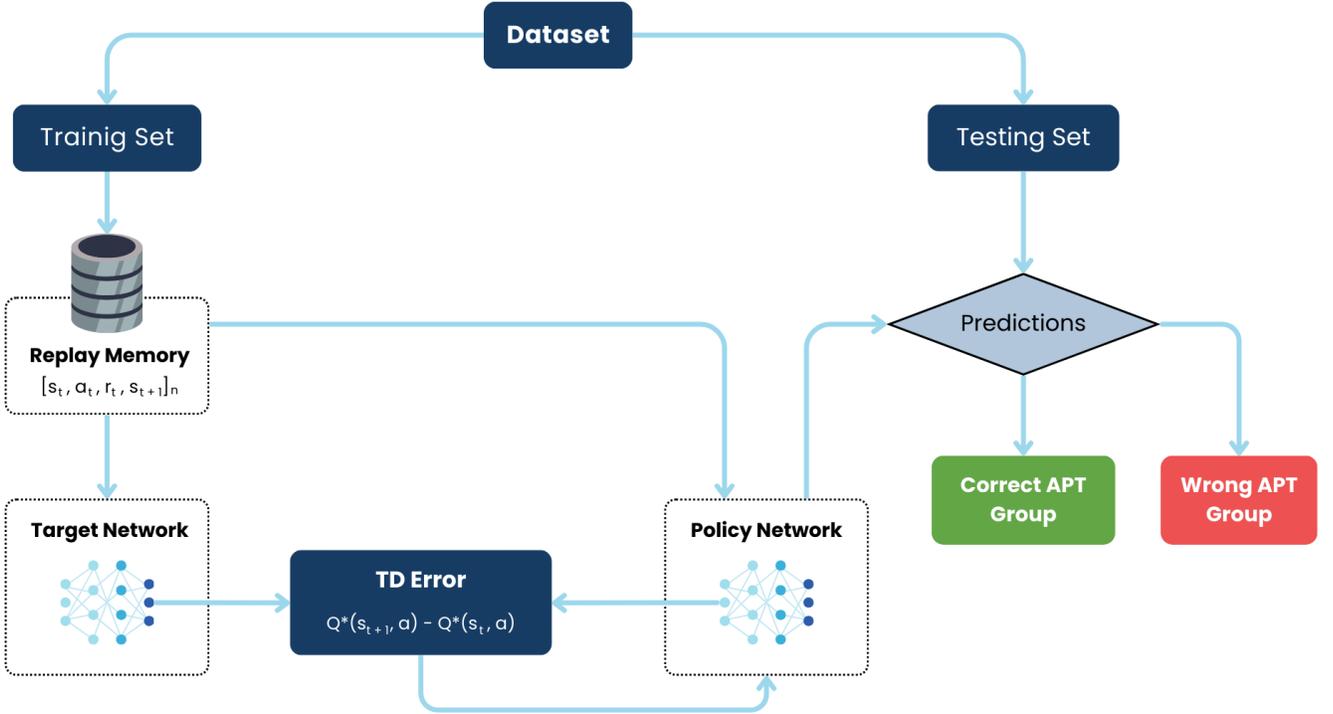

Fig. 2: DRL Model design for APT Attribution

Gymnasium offers the foundational framework necessary for designing and managing the interaction between the DRL agent and the simulated network environment, which is vital for both the training and evaluation phases. It enables the DRL model to adapt and learn efficiently from dynamic scenarios that mimic real APT attacks. [47]

## VI. RESULTS AND DISCUSSIONS

### A. Results Evaluation

The development of the DRL model for malware attribution involved extensive research, iterative coding, and numerous adjustments based on the insights gathered from predecessor models and contemporary research papers. This preparatory work was essential to establish a robust foundation for the model, ensuring it could adapt and respond effectively to the dynamic nature of malware threats. Initially, the model struggled with low accuracy levels, but through persistent adjustments to its architecture and learning algorithms, accuracy improved dramatically—from about 7% to over 73% in early iterations. By the end of the training, the model consistently reached accuracy levels near 98%, demonstrating its strong capability to accurately recognise and attribute malware activities. This upward trajectory in training accuracy is graphically represented in the Figure, which vividly illustrates the model's maturation and increasing proficiency over time.

Following the graph for training accuracy, a detailed heatmap was generated to gain insight into the model's performance across each of the APT groups, highlighting precision, recall, and F1-scores. Notably, the model demonstrated exceptional performance with 'Equation Group', achieving perfect scores across all metrics, showcasing its capability to precisely attribute actions to this well-documented APT. Similarly, 'APT 30', 'Gorgon Group', and 'Winnti' show remarkable precision and near-perfect F1-scores, reflecting the model's strength in handling sophisticated malware profiles. In contrast, 'APT 21' presents a lower recall of 93.44%, indicating a slight challenge in capturing all activities associated with this group. This variance in performance underscores areas for potential refinement, providing valuable feedback for further enhancing the model's accuracy and adaptability to diverse malware behaviours. This heatmap serves as a crucial tool for visualising the model's specific strengths and areas for improvement in malware attribution.

In parallel with the training dataset, the DRL model's accuracy on the test dataset was thoroughly evaluated, demonstrating significant improvements throughout the training process. Starting from a modest 19.30% at the initial evaluation (step 500), the model's accuracy steadily climbed to an impressive 89.27% by the final testing step (step 20000). This progression highlights the model's increasing proficiency in adapting its predictive strategies to the complexities inherent in cybersecurity data.

The analysis of the DRL model's performance on the test dataset across various APT groups reveals its accuracy through precision, recall, and F1-scores. The model excelled with 'APT 19' and 'Equation Group', achieving F1-scores of 95.8% and 99.1% respectively, underscoring its proficiency in accurately identifying and attributing their activities. In contrast, the model faced challenges with 'APT 28', where it recorded a lower precision of 62.4% and an F1-score of 68.1%, indi-

**DRL - Agent Policy Training Algorithm**

**Preconditions:** $0 \leq \gamma \leq 1$; $0.1 \leq \epsilon \leq 1$  1: **Set** $X_\rho \leftarrow s_t$
2: **Set** $X_t \leftarrow s_{t+1}$
3: **For** each episode, **repeat**:
4:     **While** $t \leq T$ **do**:
5:         **If** $\epsilon \geq 0.1$ **then**:
6:             **Select** random $a_t$     /* $\epsilon$-greedy strategy */
7:         **Else if** $\epsilon \equiv 0.1$ **then**:
8:             $Q(s_t, a_t) \leftarrow X_\rho \omega_\rho + b$
9:             **Select** $a_t : a_t \leftarrow \text{index}(\max Q(s_t, a_t))$
10:         **End If**
11:         **Observe** $r_t, s_{t+1}$
12:         **Store** experiences: $B_n \leftarrow \{(s_t, a_t, r_t, s_{t+1})_n\}$
13:         **Select** randomly $B \subset B_n$
14:         $Q(s_t, a_t) \leftarrow X_\rho \omega_\rho + b$
15:         $Q(s_{t+1}, a_{t+1}) \leftarrow X_t \omega_t + b$
16:         $Q^*(s, a) \leftarrow r_t + \gamma \max(Q(s_{t+1}, a_{t+1}))$
17:         $L(fx) \leftarrow Q^*(s, a) - Q(s, a)$
18:         $\omega_\rho \leftarrow \{\omega_\rho - \alpha \frac{dL}{d\omega_\rho}\}$     /* update weights */
19:         **If** $T \equiv f$ **then**:
20:             $X_t \omega_t + b \leftarrow X_p \omega_p + b$
21:         **End If**
22:         $t \leftarrow t + 1$
23:     **End While**
24: **End For**

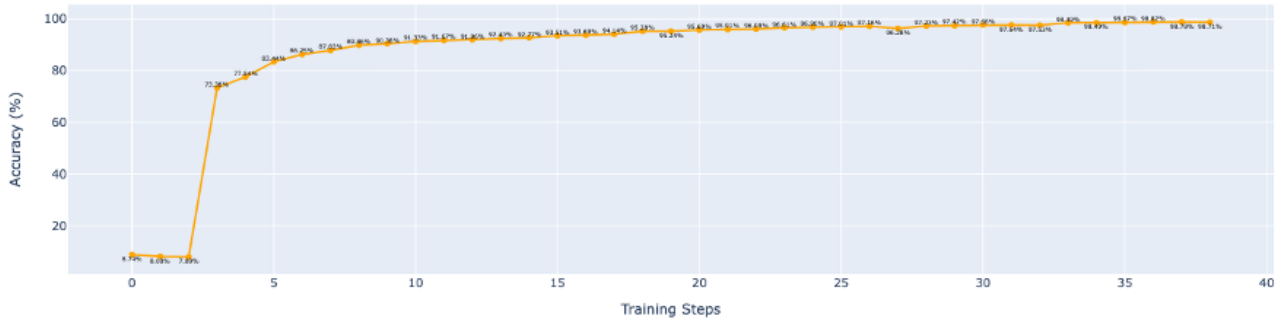

Fig. 3: Accuracy of the DRL Model for APT Attribution on Training Data
Accuracy of the DRL Model for APT Attribution on Training Data

cating difficulties in correctly detecting this group's actions. Notably, 'Energetic Bear' and 'Equation Group' displayed almost perfect precision, highlighting the model's strength in pinpointing these groups with high accuracy. Generally, the model demonstrated relatively high precision, recall, and F1-scores across most groups, reflecting its overall effectiveness in accurately attributing a diverse array of APT activities.

The overall performance of the DRL model on the test dataset can be quantified through key metrics including accuracy, precision, recall, and F1-score. The model achieved an average accuracy of 89%, indicating a high rate of correctly identifying APT-related activities. Precision averaged at 87%, suggesting that the majority of the model's predictions were relevant and accurately attributed to the correct APT groups. The recall rate of 85% reflects the model's ability to capture a substantial proportion of the actual positive cases, while the F1-score, averaging 86%, illustrates a balanced relationship between precision and recall, confirming the model's robustness in various testing scenarios.

TABLE IV: Experimental Setup for the DQN model

| Parameter | Value |
| --- | --- |
| Learning Rate Schedule | $1 \times 10^{-3} \times (0.99 \text{ step}/1000)$ |
| Policy | MlpPolicy |
| Buffer Size | 100,000 |
| Batch Size | 256 |
| Gradient Steps | 3 |
| Tau ($\tau$) | 0.005 |
| Exploration Fraction | 0.1 |
| Exploration Final Eps | 0.02 |
| Gamma ($\gamma$) | 0.99 |
| Net Architecture | [1024, 512, 512, 256] |
| Activation Function | `torch.nn.LeakyReLU` |

TABLE V: Performance Metrics for the APT Attribution Model on the Test Dataset

| Metric | Value (Average) |
| --- | --- |
| Accuracy | 89% |
| Precision | 87% |
| Recall | 85% |
| F1-score | 86% |

Multi-Layer Perceptron (MLP), and Decision Tree Classifier were implemented to represent a broad spectrum of machine learning techniques, each with its strengths and weaknesses in handling classification tasks.

TABLE VI: Comparison of test accuracy across different models, including the proposed DRL model

| Model | Test Accuracy |
| --- | --- |
| SGD | 71.47% |
| SVC | 77.41% |
| KNN | 80.21% |
| MLP | 80.65% |
| Decision Tree Classifier | 82.64% |
| **DRL Model*** | **89.27%** |

The comparative analysis revealed that the DRL model significantly outperformed the other models, achieving a test accuracy of 89.27%. In contrast, the Decision Tree Classifier, which had the next highest accuracy, reached only 82.64%. Models such as MLP and KNN also showed strong performance with accuracies of 80.65% and 80.21% respectively, while the SVC and SGD trailed with 77.41% and 71.47%. The superior performance of the DRL model underscores its advanced capability in learning from and adapting to the complex patterns of malware behaviours more effectively than traditional models. This indicates not only the robustness of the DRL approach in handling the nuances of cybersecurity threat detection but also its potential to provide more reliable and precise attributions in real-world applications.

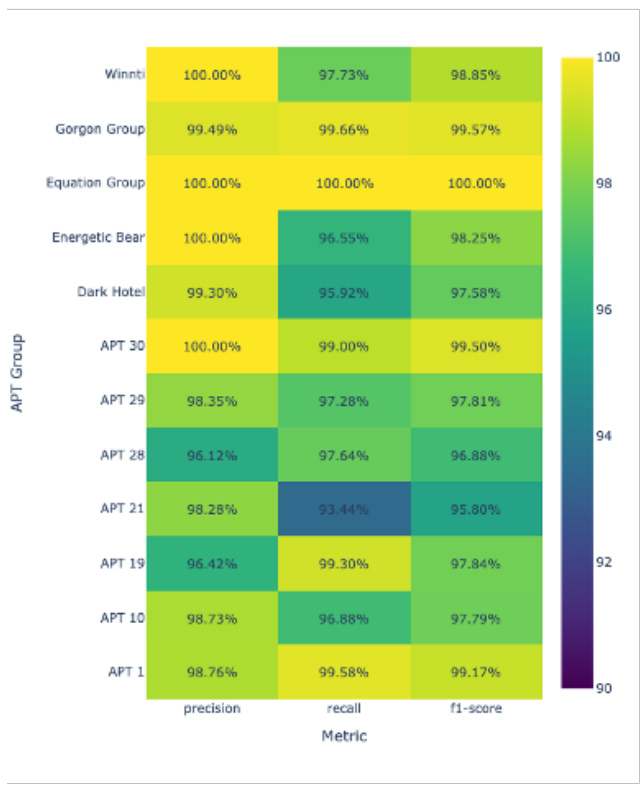

Fig. 4: Precision, Recall and F1 Score of the DRL Model for APT Attribution on each of the APT group in the Training Data

Precision, Recall and F1 Score of the DRL Model for APT Attribution on each of the APT group in the Training Data

### B. Model Comparison

Following the evaluation of the DRL model's performance, it is essential to place its achievements in the context of alternative approaches. To establish a comprehensive understanding of the DRL model's capabilities, it was benchmarked against several other machine learning models that were developed using the same dataset. This comparative analysis is pivotal as it provides a clearer picture of the DRL model's relative efficiency and accuracy in attributing malware to specific APT groups. Models such as Stochastic Gradient Descent (SGD), Support Vector Classifier (SVC), K-Nearest Neighbors (KNN),

### C. Limitations and Future Works

The study identifies several limitations in implementing the DRL model for APT attribution. One significant constraint is the high computational demand, as the DRL model requires extensive processing power and memory to handle large datasets and perform complex computations. This resource-intensive nature can limit its scalability, particularly in environments with limited hardware capabilities. Additionally, the model's effectiveness heavily depends on the availability of high-quality and diverse training data. In cybersecurity, where data is often scarce or sensitive, this dependency can restrict the model's learning potential and adaptability. The complexity of implementing and fine-tuning the DRL model also poses a challenge; its sophisticated nature requires expert knowledge in both reinforcement learning and cybersecurity, along with careful parameter adjustments to maintain optimal performance, which can be both resource-intensive and a barrier to widespread adoption.

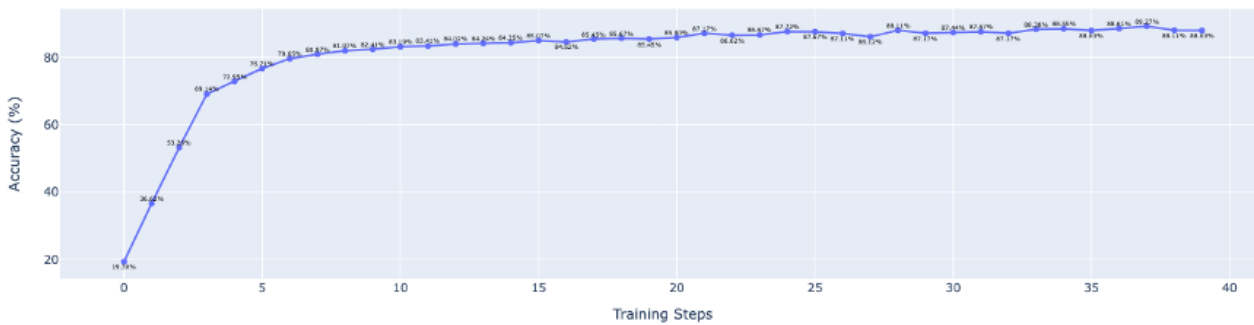

Fig. 5: Accuracy of the DRL Model for APT Attribution on Test Data

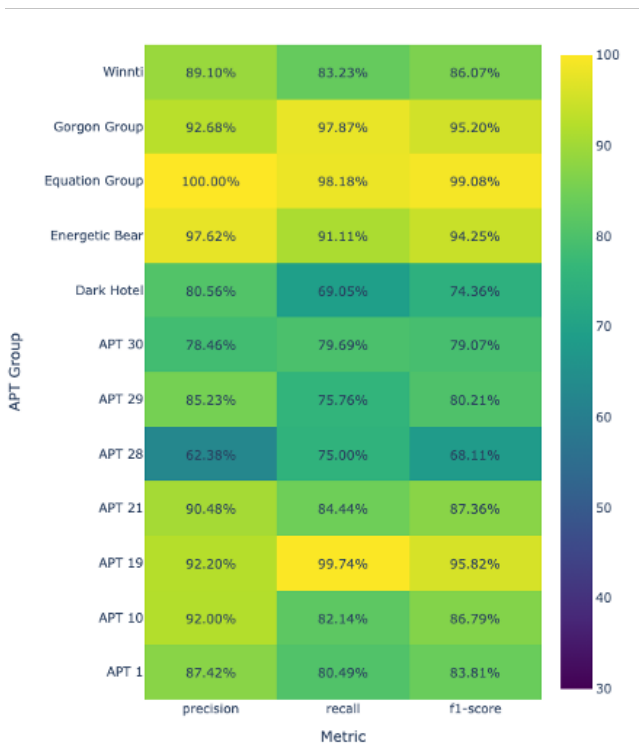

Fig. 6: Precision, Recall and F1 Score of the DRL Model for APT Attribution on each of the APT group in the Testing Data

To address these limitations, future work could focus on enhancing the computational efficiency of the DRL model by refining its architecture, employing more efficient algorithms, and utilising techniques like transfer learning and model pruning to reduce computational load without sacrificing performance. Expanding the diversity of training datasets to include a broader range of malware samples would also strengthen the model's ability to generalise and improve accuracy across different attack types. Additionally, addressing legal and ethical considerations, such as data privacy and bias, should be a priority, with guidelines developed for the ethical use of AI in cybersecurity. Finally, leveraging Large Language Models (LLMs) could further enhance DRL systems by optimising reward mechanisms and decision-making strategies. LLMs can help create more dynamic reward structures, improving the balance between exploration and exploitation and ultimately boosting the model's capacity to detect and respond to complex security threats [48].

## VII. CONCLUSION

This research demonstrates the significant advancements in the application of Deep Reinforcement Learning (DRL) for attributing Advanced Persistent Threat (APT) groups, using a detailed dataset of over 3,500 malware samples across 12 distinct APT groups. The DRL model showcased its capabilities by significantly outperforming traditional machine learning approaches such as Stochastic Gradient Descent (SGD), Support Vector Classifier (SVC), K-Nearest Neighbours (KNN), Multi-Layer Perceptron (MLP), and Decision Tree Classifiers. With a remarkable test accuracy of 89.27%, the DRL model stands out, not only for its high precision in malware attribution but also for its adaptability to the complex and evolving landscape of cyber threats. By applying DRL, organisations can enhance their threat intelligence capabilities, allowing for more nuanced understanding and preemptive actions against APTs. This study's findings underscore the potential of DRL in enhancing cybersecurity operations by providing rapid and accurate threat attribution, paving the way for further research on its applicability across more diverse datasets and optimising its computational efficiency for broader use in real-world scenarios.

**Research Ethics**: This study was deemed exempt from ethics approval as it did not involve human or animal subjects.

**Code and Data**: The code and datasets used and generated during this research are made publicly available at https://github.com/crypticsy/APTAttribution

## REFERENCES

[1] kasperskyLab. The power of threat attribution: Challenges and benefits of cyberthreat attribution — kaspersky.com. https://media.kaspersky.com/en/business-security/enterprise/threat-attribution-engine-whitepaper.pdf, 2020. [Accessed 12-05-2024].

[2] Weijie Han, Jingfeng Xue, Yong Wang, Zhenyan Liu, and Zixiao Kong. Malinsight: A systematic profiling based malware detection framework. *Journal of Network and Computer Applications*, 125:236–250, 2019.

[3] BinHui Tang, JunFeng Wang, Zhongkun Yu, Bohan Chen, Wenhan Ge, Jian Yu, and TingTing Lu. Advanced persistent threat intelligent profiling technique: A survey. *Computers and Electrical Engineering*, 103:108261, 2022.

[4] Marie Baezner and Patrice Robin. Stuxnet. Report 4, Zurich, 2017-10-18.

[5] Edward Kost. What is an Advanced Persistent Threat (APT)? | UpGuard — upguard.com. https://www.upguard.com/blog/what-is-an-advanced-persistent-threat, 2023. [Accessed 04-08-2024].

[6] Kazeem Saheed and Shagufta Henna. Deep reinforcement learning for advanced persistent threat detection in wireless networks. In *2023 31st Irish Conference on Artificial Intelligence and Cognitive Science (AICS)*, pages 1–6, 2023.

[7] The Hacker News. 3 Ransomware Group Newcomers to Watch in 2024 — thehackernews.com. https://thehackernews.com/2024/01/3-ransomware-group-newcomers-to-watch.html, 2024. [Accessed 06-08-2024].

[8] Djallel Hamouda, Mohamed Amine Ferrag, Nadjette Benhamida, Hamid Seridi, and Mohamed Chahine Ghanem. Revolutionizing intrusion detection in industrial iot with distributed learning and deep generative techniques. *Internet of Things*, 26:101149, 2024.

[9] Robin Buchta, George Gkoktsis, Felix Heine, and Carsten Kleiner. Advanced persistent threat attack detection systems: A review of approaches, challenges, and trends. *Digital Threats*, September 2024. Just Accepted.

[10] Muhammad Raza. What Are TTPs? Tactics, Techniques & Procedures Explained | Splunk — splunk.com. https://www.splunk.com/en_us/blog/learn/ttp-tactics-techniques-procedures.html, 2023. [Accessed 10-08-2024].

[11] Joaquin Matamis. Advancing Accountability in Cyberspace • Stimson Center — stimson.org. https://www.stimson.org/2024/advancing-accountability-in-cyberspace/. [Accessed 05-09-2024].

[12] Tinshu Sasi, Arash Habibi Lashkari, Rongxing Lu, Pulei Xiong, and Shahrear Iqbal. A comprehensive survey on iot attacks: Taxonomy, detection mechanisms and challenges. *Journal of Information and Intelligence*, 2023.

[13] Dipo Dunsin, Mohamed Chahine Ghanem, Karim Ouazzane, and Vassil Vassilev. Reinforcement learning for an efficient and effective malware investigation during cyber incident response. *arXiv preprint arXiv:2408.01999*, 2024.

[14] Eduardo C. Garrido Merchán. Why deep reinforcement learning is going to be the next big deal in AI — eduardogarrido90. https://medium.com/, 2023. [Accessed 10-08-2024].

[15] Sang Ho Oh, Jeongyoon Kim, Jae Hoon Nah, and Jongyoul Park. Employing deep reinforcement learning to cyber-attack simulation for enhancing cybersecurity. *Electronics*, 13(3), 2024.

[16] Mangadevi Atti and Manas Yogi. Application of deep reinforcement learning (drl) for malware detection. *International Journal of Information technology and Computer Engineering*, 4:23–35, 04 2024.

[17] Shui Yu, Guofei Gu, Ahmed Barnawi, Song Guo, and Ivan Stojmenovic. Malware propagation in large-scale networks. *IEEE Transactions on Knowledge and Data Engineering*, 27(1):170–179, 2015.

[18] Mohamed C Ghanem, Thomas M Chen, Mohamed A Ferrag, and Mohyi E Kettouche. Esascf: expertise extraction, generalization and reply framework for optimized automation of network security compliance. *IEEE Access*, 2023.

[19] Mohamad Fadli Zolkipli and Aman Jantan. Malware behavior analysis: Learning and understanding current malware threats. In *2010 Second International Conference on Network Applications, Protocols and Services*, pages 218–221, 2010.

[20] Nana Kwame Gyamfi, Nikolaj Goranin, Dainius Ceponis, and Habil Antanas Čenys. Automated system-level malware detection using machine learning: A comprehensive review. *Applied Sciences*, 13(21), 2023.

[21] Chaoxian Wei, Qiang Li, Dong Guo, Xiangyu Meng, and Angel M. Del Rey. Toward identifying apt malware through api system calls. *Sec. and Commun. Netw.*, 2021, January 2021.

[22] Mohammed Ashfaaq M Farzaan, Mohamed Chahine Ghanem, Ayman El-Hajjar, and Deepthi N Ratnayake. Ai-enabled system for efficient and effective cyber incident detection and response in cloud environments. *arXiv preprint arXiv:2404.05602*, 2024.

[23] Ishai Rosenberg, Guillaume Sicard, and Eli (Omid) David. End-to-end deep neural networks and transfer learning for automatic analysis of nation-state malware. *Entropy*, 20(5), 2018.

[24] Md Hasan, Muhammad Usama Islam, and Jasim Uddin. Advanced persistent threat identification with boosting and explainable ai. *SN Computer Science*, 4, 03 2023.

[25] Binhui Tang, Peiyun Leng, Xuxiang Shen, and Yuhang Wei. Deep learning-based apt malware and variants detection with attribution analysis. pages 996–1003, 08 2023.

[26] Elijah Snow, Mahbubul Alam, Alexander Glandon, and Khan Iftekharuddin. End-to-end multimodel deep learning for malware classification. In *2020 International Joint Conference on Neural Networks (IJCNN)*, pages 1–7, 2020.

[27] Gil Shenderovitz and Nir Nissim. Bon-apt: Detection, attribution, and explainability of apt malware using temporal segmentation of api calls. *Computers & Security*, 142:103862, 2024.

[28] Mohamed Chahine Ghanem, Patrick Mulvihill, Karim Ouazzane, Ramzi Djemai, and Dipo Dunsin. D2wfp: a novel protocol for forensically identifying, extracting, and analysing deep and dark web browsing activities. *Journal of Cybersecurity and Privacy*, 3(4):808–829, 2023.

[29] Jian Zhang, Shengquan Liu, and Zhihua Liu. Attribution classification method of APT malware based on multi-feature fusion. *PLoS One*, 19(6):e0304066, June 2024.

[30] Shudong Li, Qianqing Zhang, Xiaobo Wu, Weihong Han, and Zhihong Tian. Attribution classification method of apt malware in iot using machine learning techniques. *Security and Communication Networks*, 2021:1–12, 09 2021.

[31] Dipo Dunsin, Mohamed Chahine Ghanem, Karim Ouazzane, and Vassil Vassilev. Reinforcement learning for an efficient and effective malware investigation during cyber incident response. *arXiv preprint arXiv:2408.01999*, 2024.

[32] Cho Do Xuan and Nguyen Hoa Cuong. A novel approach for APT attack detection based on feature intelligent extraction and representation learning. *PLoS One*, 19(6):e0305618, June 2024.

[33] Pawel Ladosz, Lilian Weng, Minwoo Kim, and Hyondong Oh. Exploration in deep reinforcement learning: A survey. *Information Fusion*, 85:1–22, 2022.

[34] Kai Arulkumaran, Marc Peter Deisenroth, Miles Brundage, and Anil Anthony Bharath. Deep reinforcement learning: A brief survey. *IEEE Signal Processing Magazine*, 34(6):26–38, 2017.

[35] Manasa Vemuri. Malware Analysis using Cuckoo Sandbox — easypeazyo14. https://medium.com/@easypeazyo14/malware-analysis-using-cuckoo-sandbox-756616e6e85e, 2020. [Accessed 11-10-2024].

[36] cuckoosandbox. Cuckoo Sandbox - Automated Malware Analysis — cuckoosandbox.org. https://cuckoosandbox.org/, 2024. [Accessed 11-10-2024].

[37] Samet. What is Virus Total? — sametyorulmaz777. https://medium.com/@sametyorulmaz777/what-is-virus-total-70c64b7c5e95, 2022. [Accessed 11-10-2024].

[38] VirusTotal. VirusTotal — virustotal.com. https://www.virustotal.com/gui/intelligence-overview, 2024. [Accessed 11-10-2024].

[39] Mohamed Chahine Ghanem and Deepthi N Ratnayake. Enhancing wpa2-psk four-way handshaking after re-authentication to deal with de-authentication followed by brute-force attack a novel re-authentication protocol. In *2016 International Conference On Cyber Situational Awareness, Data Analytics And Assessment (CyberSA)*, pages 1–7. IEEE, 2016.

[40] CyberResearch. GitHub - cyber-research/APTMalware: APT Malware Dataset Containing over 3,500 State-Sponsored Malware Samples — github.com. https://github.com/cyber-research/APTMalware, 2019. [Accessed 05-10-2024].

[41] Manuel Lopez-Martin, Belen Carro, and Antonio Sanchez-Esguevillas. Application of deep reinforcement learning to intrusion detection for supervised problems. *Expert Systems with Applications*, 141:112963, 2020.

[42] Xiaoguang Li. Create custom openai gym environment for deep reinforcement learning (drl4t-04). https://lixiaoguang.medium.com/, 2023. [Accessed 15-08-2024].

[43] Juan C Olamendy. Understanding ReLU, LeakyReLU, and PReLU: A Comprehensive Guide — juanc.olamendy. https://medium.com/@juanc.olamendy/understanding-relu-leakyrelu-and-prelu-a-comprehensive-guide-20f2775d3d64, 2023. [Accessed 18-08-2024].

[44] scikitLearn. MinMaxScaler — scikit-learn.org. https://scikit-learn.org/stable/modules/generated/sklearn.preprocessing.MinMaxScaler.html, 2024. [Accessed 16-08-2024].

[45] Neil Fox. Cuckoo Sandbox Overview — varonis.com. https://www.varonis.com/blog/cuckoo-sandbox, 2023. [Accessed 17-08-2024].

[46] Stable Baselines 3. Stable-Baselines3 Docs - Reliable Reinforcement Learning Implementations & x2014; Stable Baselines3 2.4.0a9 documentation — stable-baselines3.readthedocs.io. https://stable-baselines3.readthedocs.io/en/master/, 2024. [Accessed 20-08-2024].

[47] Farama Foundation. Gymnasium Documentation — gymnasium.farama.org. https://gymnasium.farama.org/index.html, 2023. [Accessed 19-08-2024].

[48] Fatma Yasmine Loumachi and Mohamed Chahine Ghanem. Advancing cyber incident timeline analysis through rule based ai and large language models. *arXiv preprint arXiv:2409.02572*, 2024.